# An Algebraic Approach to Fast Estimation of the Threshold Voltage of Junctionless Double Gate MOSFETs Using the Gram–Schmidt Method

Mohsen Annabestani, Mahshid Nasserian, Fatemeh Hasanzadeh, Mohammad Taherzadeh-Sani*, and Alireza Hassanzadeh

*Abstract*— The effect of decreasing Drain-Induced Barrier Lowering (DIBL) is one of the non-desirable short-channel effects in the MOSFETs family, which causes the threshold voltage of the transistor to be reduced by increasing the voltage of the drain. This effect makes it impossible for circuit designers to consider $V_T$ as a constant value, and hence, it is necessary to calculate $V_T$ as a function of the drain voltage. Therefore, to consider the effect of DIBL in the design of integrated circuits, a large computational burden is imposed on the system, which slows down the simulation process in circuit-level simulators, particularly when a large number of transistors are to be simulated. Accordingly, in this paper, a multiple input single output (MISO) Nonlinear Autoregressive (N-AR) model using the Gram-Schmidt orthogonalization approach is proposed, that calculates the threshold voltage of the new generation of MOSFETs, i.e., Junctionless Double-Gate MOSFETs (JL-DG-MOSFETs), with a high precision and a significant speed-up in the computational procedure of the model. It is shown that, on average, the proposed numerical method is 313 times faster than the state-of-the-art analytical model. The calculated percentage of normalized mean square error between the proposed model and analytical one is 0.435% on average, showing that the proposed approach can be a fast and accurate candidate for replacing the analytical modeling.

## I. INTRODUCTION

Drain-Induced Barrier Lowering (DIBL) is one of the short channel effects in MOSFETs, which causes the threshold voltage of the transistor to be reduced by increasing the drain voltage of the transistor. Hence, the threshold voltage cannot be considered as a constant value. Researchers have proposed double-gate MOSFETs (DG-MOSFETs) to reduce this effect, as well as to reduce other short channel effects. Other advantages of DG-MOSFETs include higher transconductance and better control of charge carriers in the channel [1, 2]. Recently, a new type of DG-MOSFETs have been introduced that do not have a junction, and they are referred to as Junctionless DG-MOSFETs (JL-DG-MOSFET) [3-5]. JL-DG-MOSFETs have the same structure of double gate MOSFETs, but the transistor's body is doped through a channel between source and drain, as shown in Fig. 1. This doped structure has special advantages from the manufacturing process point of view and also improves the performance of the device [6, 7]. The advantages of the JL-DG-MOSFET compared to common MOSFET are a low subthreshold slope, high $I_{ON}/I_{OFF}$ ratio, and less leakage current. Hence, based on these features, JL-DG-transistors are beneficial for applications such as RF analog circuits and memory chips.

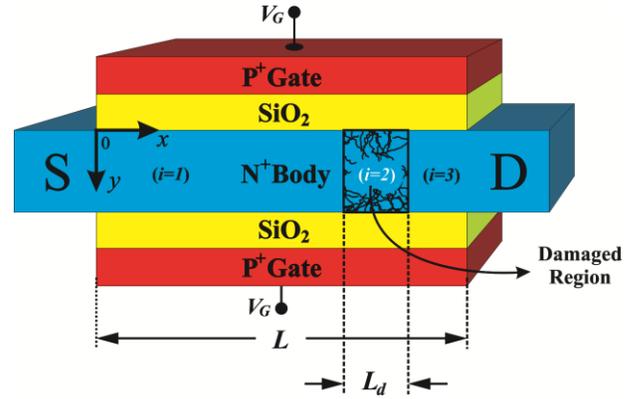

**Fig. 1.** Side view of a JL-DG-MOSFET. The letter *i* represents to a damaged/undamaged region, $i=2$ refers to the damaged region and $i=1,3$ represent the undamaged regions. Also, $L$ and $L_d$ are the channel length and damaged channel length, respectively.

In a recent work, the effect of non-uniform doping on the characteristics of symmetric JL-DG-MOSFETs has been investigated by employing an empirical modeling approach [8]. Various works have focused on the modeling of JL-DG-transistors based on analytical formulation. Authors in [9] investigate the characteristics of an inverter made by JL-DG-transistors using a mixed-mode simulation. Moreover, in [10], the effect of temperature on the DC and AC performances of JL-DG-transistors is investigated. In [11], a charge-based compact model for JL-DG-transistors is developed to investigate the DC and the quasi-static characteristics of the transistor technology. Researchers in [12] studies electron transport in Si JL-DG-transistors using multisubband Monte Carlo method, which shows that these transistors can decrease the mobility reduction due to surface roughness scattering and enhance capability of the current driving.

Moreover, other features of short channel symmetric and asymmetric JL-DG-transistors to be used in different applications are exploited in [13-18]. For example, to find the I-V specifications of cylindrical nanowire junctionless MOSFETs, an analytical quantum-mechanical model is developed in [13] using the Landauer formalism. The work in [13] considers the relevant quantum effects resulting from the cylindrical carried confinement, and it handles equations for the device features. In [17], using an analytical model approach, authors propose the idea of the Gaussian-like doping inside the channel of JL-DG-transistors. This work employs the approximation by replacing the nonanalytical Gaussian function with the Gaussian-like function for traditional DG-MOSFET.

In spite of overall better performance, JL-DG-transistors still have the disadvantage of the Hot Carrier Effect (HCE). Various analytical models have been developed to calculate the threshold voltage, which partly considers the problem of the hot carrier's effect [19, 20]. For example, the aim of authors in [6] is to present an analytical model for calculating threshold voltages of a JL-DG-MOSFET [6]. To explore the effect of localized interface charges on depletion width in these MOSFETs, the following 2-D Poisson's equation is used:

$$\frac{\partial^2 \phi_s}{\partial x^2} + \frac{\partial^2 a_1}{\partial x^2} y_d + \frac{\partial^2 a_2}{\partial x^2} y_d^2 + 2a_2 = -\frac{qN_{sub}}{\varepsilon_{si}} \quad (1)$$

Where $\phi_s$ is the surface potential, $N_{sub}$ is the body doping concentration, $y_d$ is the depletion width along the y-axis. Also, $a_1$ and $a_2$ are parameters expressed in terms of $y_d$ and $\phi_s$ by considering the electric flux continuity and using Gauss's law in boundary conditions. By solving (1) and then applying some physical and mathematical efforts, a well-defined equation is obtained for the threshold voltage ($V_T$) [6]:

$$V_T = \max_{i=1}^{3}(V_{Ti}) \quad (2)$$

where

$$V_{Ti} = V_{T0} - 2\sqrt{b_i c_i} - \frac{qN_f}{C_{ox}} \delta(i-2) \quad (3)$$

In (3), $N_f$ is the trap charge density and $\delta(i-2)$ is a shifted delta function, which is equal to 1 for $i = 2$, and it is equal to 0 for $i = 1, 3$. Also, as mentioned in the [6], $b_i c_i$ is a quadratic equation of $V_{Ti}$. Furthermore, $V_{T0}$ is defined as

$$V_{T0} = V_{FB0} + V_C - \frac{qN_{sub}t_{si}}{2C_{ox}} - \frac{qN_{sub}t_{si}^2}{8C_{ox}} \quad (4)$$

where $V_{FB0}$, $V_C$ and $t_{si}$ are the flat-band voltages for the undamaged device, the central voltage, and the silicon thickness, respectively. However, the problem with analytical models, including [6], is that the DIBL effect is still present in them, and therefore, the threshold voltage must be calculated as a function of the drain voltage, frequently at each time. This issue results in increasing the computational time overhead and reducing the design efficiency, especially in integrated circuits with a large number of transistors. Therefore, in this paper, a model is proposed that approximates the performance of the analytical models with a desirable accuracy and a much higher speed. This method can significantly increase the efficiency and speed of the simulation of the circuits. The proposed Multiple Input Single Output (MISO) Nonlinear Autoregressive (N-AR) model is based on parameters of the analytical model of [6]. In the proposed model, the Gram-Schmidt orthogonalization method has been used to make a large and significant speed-up in the computational procedure of model. It will be demonstrated that the proposed model approximates the analytical model in [6] at a much shorter time with the desired accuracy.

The rest of paper is organized as follows. The Proposed N-AR model is described in Section II. In Section III, the method of Gram-Schmidt orthogonalization is presented. The simulation results of the proposed method for estimating the threshold voltage of JL-DG-MOSFET is provided in Section IV.

## II. THE PROPOSED N-AR MODEL

In this paper, the proposed model is a truncated version of the Nonlinear Auto-Regressive with eXogenous inputs (NARX) model. This model is truncated since it does not have exogenous inputs, and hence, here, it is named Nonlinear Autoregressive (N-AR) model. A wide range of the nonlinear dynamic system can be described with the input $u$ and the output $y$ in the NARX structure using the following equation [21-25]:

$$y(k+1) = f(R(k))$$
$$R(k) = [y(k),...,y(k-n_y+1),u(k),...,u(k-n_u+1)]^T \quad (5)$$

where $y(k+1)$ refers to the predicted output at the future time $(k+1)$ and $R(k)$ is a regressor vector that includes finite numbers of the past inputs and outputs. Dynamic order of this system is determined by the number of $n_u$ and $n_y$ lags. Block diagram of a NARX system is shown in Fig. 2 [23].

In general, it can be shown that $y(k)$ can be derived from a combination of inputs and theirs lags. Hence, we can find a nonlinear function such as $f(.)$ to determine the relationship between the inputs, their lags and the output. In this paper, we want to model the static behavior of threshold voltage of a JL-DG-MOSFET, and hence, we don't need to capture the memory of the system, and we can remove all of the lag terms from the regressor vector, $R(k)$. Generally, a NARX structure can capture the memory of the system in the dynamic models, but also it is summarizable to a static model by removing the lag terms in its regressor vector. Here for JL-DG-MOSFET let's assume N-AR model as a MISO system that has $m$ inputs and one output, where the system's inputs are $u_1(k)$ to $u_m(k)$, and the output is $y(k)$. Accordingly:

$$y(k) = f(u_1(k), u_2(k), \cdots, u_m(k)) \quad (6)$$

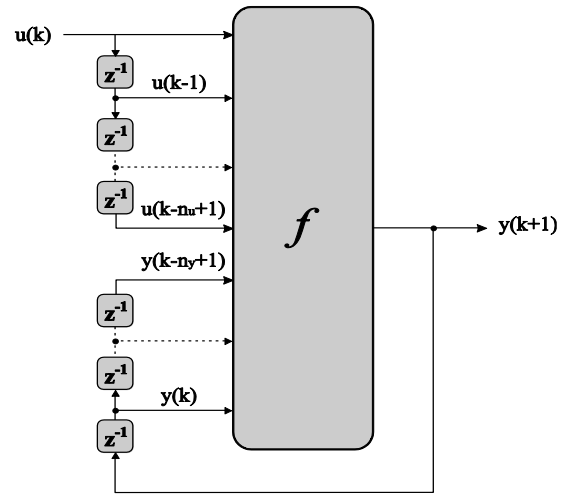

**Fig. 2.** Block diagram of a NARX system [23].

The purpose of this model is to find unknown function correctly $f(.)$. There are various techniques to find $f(.)$, including the neural networks, fuzzy neural systems, wavelet transform, spline functions, and polynomial functions, etc. Here, one of the simplest types of the function $f(.)$ has been selected, which is the polynomial function [26]. In order to model JL-DG-MOSFET, the proposed N-AR model creates a nonlinear mapping between input parameters including channel length ($L$), damaged channel length ($L_d$), silicon thickness ($t_{si}$), the thickness of the oxide ($t_{ox}$), the central voltage ($V_C$), the drain voltage ($V_D$), to generate the output parameter, the threshold voltage ($V_T$), which will be described in Section IV in details.

To rewrite equation (6) based on the polynomial function, the following equation is used:

$$y(k) = \theta_0 + \sum_{i_1=1}^{m} \theta_{i_1} u_{i_1}(k) + \sum_{i_1=1}^{m}\sum_{i_2=1}^{m} \theta_{i_1 i_2} u_{i_1}(k) u_{i_2}(k) + \cdots$$
$$\cdots + \sum_{i_1=1}^{m}\cdots\sum_{i_n=1}^{m} \theta_{i_1\cdots i_n} u_{i_1}(k) u_{i_2}(k) \cdots u_{i_n}(k) \quad (7)$$

Each of the summation terms in (7) can be considered as a linear regression equation [26] and can be rewritten as:

$$z(k) = \sum_{i=1}^{M} P_i(k) \theta_i + \xi(k) \quad t = 1, 2, ..., N \quad (8)$$

where $N$ is the length of the sampled data, $P_i(k)$ includes monomials $u_1(k)$ to $u_m(k)$, $\xi(k)$ is the modeling errors, and $\theta_i$ is the vector of unknown parameters of the model. Also, $M$ is equal to all $k$ combinations of the set $m$ without repetition such that $m$ is the number of inputs and $k = 1, 2, ..., m$. Here, $M$ can be obtained as:

$$M = \sum_{k=1}^{m} \frac{m!}{k!(m-k)!} \quad (9)$$

Let's consider (8) in the matrix form:

$$z_{N\times 1} = P_{N\times M} \Theta_{M\times 1} + \Xi_{N\times 1} \quad (10)$$

where based on the polynomial function of (7), the matrix $P_{N\times M}$ will be defined as (11).

$$P_{N\times M} = \begin{bmatrix} u_1(1) & u_2(1) & \cdots & u_m(1) & u_1(1)\times u_2(1) & \cdots & u_{m-1}(1)\times u_m(1) & \cdots & u_1(1)\times u_2(1)\times u_3(1)\times\cdots\times u_m(1) \\ u_1(2) & u_2(2) & \cdots & u_m(2) & u_1(2)\times u_2(2) & \cdots & u_{m-1}(2)\times u_m(2) & \cdots & u_1(2)\times u_2(2)\times u_3(2)\times\cdots\times u_m(2) \\ \vdots & \vdots & \cdots & \vdots & \vdots & \cdots & \vdots & \cdots & \vdots \\ \vdots & \vdots & \cdots & \vdots & \vdots & \cdots & \vdots & \cdots & \vdots \\ u_1(N) & u_2(N) & \cdots & u_m(N) & u_1(N)\times u_2(N) & \cdots & u_{m-1}(N)\times u_m(N) & \cdots & u_1(N)\times u_2(N)\times u_3(N)\times\cdots\times u_m(N) \end{bmatrix} \quad (11)$$

The goal in here is to find the vector of unknown parameters, i.e., the vector $\Theta$. Since there is always some error associated with any modeling, the vector $\Theta$ cannot be determined. Due to the error of the model, the values inside the vector also have a negligible amount of error. Hence, the resulting vector is called approximated $\Theta$, or $\hat{\Theta}$. To estimate the vector $\hat{\Theta}$, the expression $\|z - P\Theta\|$ should be minimized. According to the least mean squares (LMS) method, to solve this, the conditions of the normal equation (12) must be satisfied.

$$P^T P \Theta = P^T z \quad (12)$$

Accordingly, the vector $\hat{\Theta}$ is calculated as

$$\hat{\Theta} = (P^T P)^{-1} P^T z \quad (13)$$

So far, the usual calculation process for finding the vector $\hat{\Theta}$ is described. However, here, the matrix $P_{N\times M}$ is very large, and this issue increases the computational load of the model greatly, especially in the case of matrix inverting. In the following, an algebraic method is proposed to reduce the computational complexity of the utilized model.

### III. THE METHOD OF GRAM-SCHMIDT ORTHOGONALIZATION TO ESTIMATE VECTOR $\hat{\Theta}$

In this section, we start by dividing the matrix $P$ into the product of two matrixes, as:

$$P = WA \quad (14)$$

where $A$ is a $M \times M$ upper-triangular matrix and $W$ is an $N \times M$ matrix whose columns are orthogonal. Hence, $W^T W$ constructs a diagonal matrix called $D$.

$$A = \begin{bmatrix} 1 & \sigma_{12} & \sigma_{13} & \cdots & \sigma_{1M} \\ & 1 & \sigma_{23} & \cdots & \sigma_{2M} \\ & & \ddots & \ddots & \vdots \\ & & & 1 & \sigma_{(M-1)M} \\ & & & & 1 \end{bmatrix} \quad (15)$$

$$W = [w_1, \cdots, w_M] \quad (16)$$

$$D = W^T W \quad (17)$$

The procedure of Gram-Schmidt orthogonalization is shown in (18).

$$\left.\begin{array}{l} w_1 = p_1 \\ \sigma_{ir} = \dfrac{\langle w_i, p_r \rangle}{\langle w_i, w_i \rangle} \\ w_r = p_r - \sum_{i=1}^{r-1} \sigma_{ir} w_i \end{array}\right\} \quad 1 \leq i \leq r \quad (18)$$

The expression $\langle w_i, p_r \rangle$ denotes the inner multiplication $p_r$ and $w_i$, which is defined as follows:

$$\langle w_i, p_r \rangle = w_i^T p_r = \sum_{i=1}^{M} w_i(k) p_r(k) \quad (19)$$

It is defined here:

$$g = D^{-1} W^T z \quad (20)$$

$$g_i = \dfrac{\langle w_i, z \rangle}{\langle w_i, w_i \rangle} \quad i = 1, 2, \cdots, M \quad (21)$$

Finally, the vector $\hat{\Theta}$ is obtained as:

$$A\Theta = g \Rightarrow \hat{\Theta} = A^{-1} g \quad (22)$$

The decomposition of the Gram-Schmidt orthogonalization method eliminates the inverting process of matrix $P$, and the model should only deal with the simple process of inverting the matrix $A$. Consequently, the computational burden is greatly reduced. This simplicity is due to the lattice structure of the matrix $A$. For example, it can be shown that the inverse of an upper-triangular matrix is an upper-triangular matrix, and the determinant of an upper-triangular matrix is obtained only by calculating the multiplication of its diagonal elements. Also, the diagonal elements of the inverse of the matrix $A$ are the inverse of its diagonal elements [27].

## IV. SIMULATION RESULTS OF THE PROPOSED METHOD FOR ESTIMATING JL-DG-MOSFET THRESHOLD VOLTAGE

As mentioned before, according to the analytical model presented in [6], it is needed to use the parameters $L$, $L_d$, $t_{si}$, $t_{ox}$, $V_C$ and $V_D$ to calculate the threshold voltage, $V_T$. By using (6) to solve the threshold voltage estimation, the following relation is resulted:

$$V_T(k) = f(L(k), L_d(k), t_{ox}(k), t_{si}(k), V_C(k), V_D(k)) \quad (23)$$

As described in the previous section, the vector $\hat{\Theta}$ parameters can be calculated and then the $V_T$ will be approximated. For a better understanding, we obtain the matrix $P$ for this problem, which is given in (24).

$$P_{N \times M} = \begin{bmatrix} L(1) & L_d(1) & \cdots & V_D(1) & L(1) \times L_d(1) & \cdots & V_C(1) \times V_D(1) & \cdots & L(1) \times L_d(1) \times t_{si}(1) \times \cdots \times V_D(1) \\ L(2) & L_d(2) & \cdots & V_D(2) & L(2) \times L_d(2) & \cdots & V_C(2) \times V_D(2) & \cdots & L(2) \times L_d(2) \times t_{si}(2) \times \cdots \times V_D(2) \\ \vdots & \vdots & \cdots & \vdots & \vdots & \cdots & \vdots & \cdots & \vdots \\ \vdots & \vdots & \cdots & \vdots & \vdots & \cdots & \vdots & \cdots & \vdots \\ L(N) & L_d(N) & \cdots & V_D(N) & L(N) \times L_d(N) & \cdots & V_C(N) \times V_D(N) & \cdots & L(N) \times L_d(N) \times t_{si}(N) \times \cdots \times V_D(N) \end{bmatrix} \quad (24)$$

Typically, there are three basic stages in a system identification routine including data creation, model determination, and validation [28, 29]. Model determination stage was explained in the previous sections. In this section, based on simulated data obtained from the analytical method of [6], the validation stage is investigated. Based on Eq (23) the proposed model has 6 inputs, and if we want to involve all values of all inputs in training procedure we need to burden a very high computational load and it is not acceptable for our model, hence to solve this problem the proposed model is trained by a discrete sequence of the random numbers sampled from the proper range of inputs. Testing results will show that this training approach using sampled random numbers is a proper solution. For model validation, the trained model has been tested by the discrete sampled sequence of chirp, sinusoidal and quasi-triangular (Q-Triangular) datasets (Figs. 3-8). Here the main inputs are $V_D(k)$, and $V_C(k)$ the calculated $V_T$ (in Volt) using the method of [6] is output. The constant values ($t_{ox}$, $t_{si}$, $L$, and $L_d$) have also been swept to cover more variability in the model validation and making sure the proposed model can work under the verity of variations. In this study $t_{ox}$, $t_{si}$, $L$, and $L_d$ have been swept in the /ranges of [1, 4]$^{nm}$, [5, 15]$^{nm}$, [20, 40]$^{nm}$, and [0, 20]$^{nm}$, respectively.

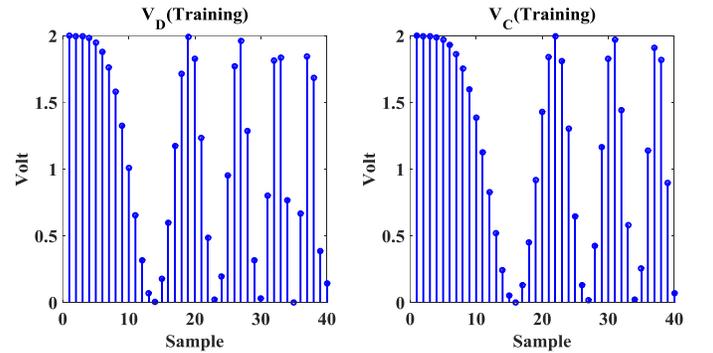

**Fig. 3.** The discrete sequence of the chirp signal as the inputs samples.

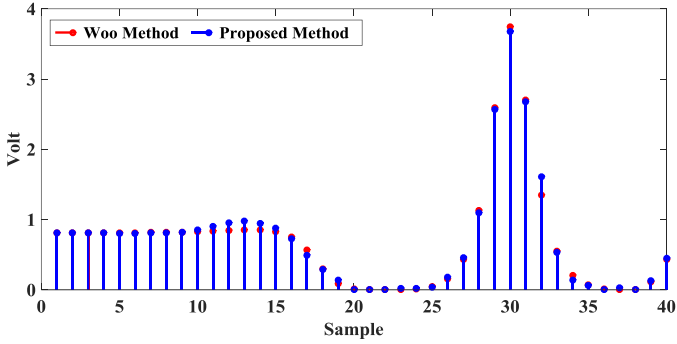

**Fig. 4.** The output ($V_T$) of the proposed model compared to the method of [6](Woo's model) for the discrete Chirp testing data set.

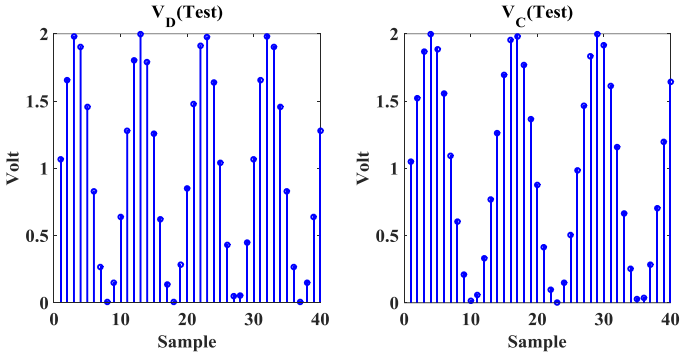

**Fig. 5.** The discrete sequence of the sinusoidal signal as the input samples.

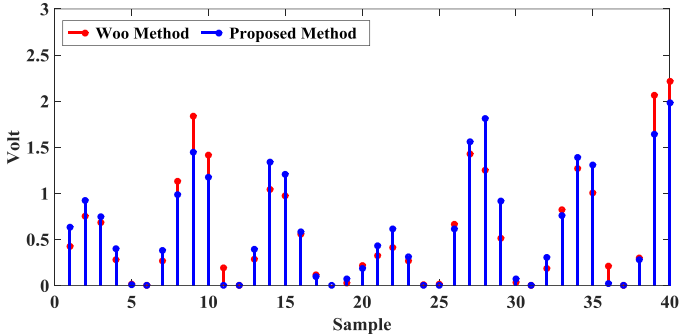

**Fig. 6.** The output ($V_T$) of the proposed model compared to the method of [6] (Woo's model) for the discrete sinusoidal testing data set.

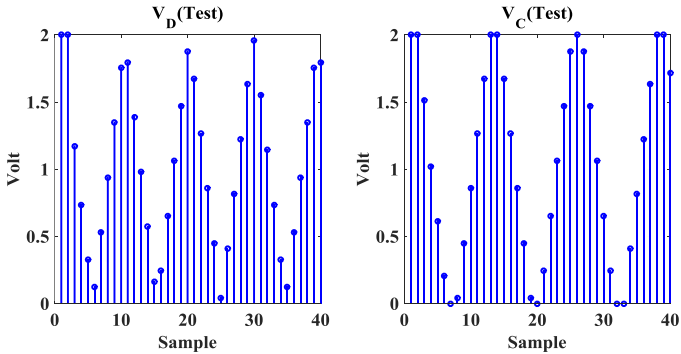

**Fig. 7.** The discrete sequence of the quasi-triangular (Q-triangular) signal as input samples.

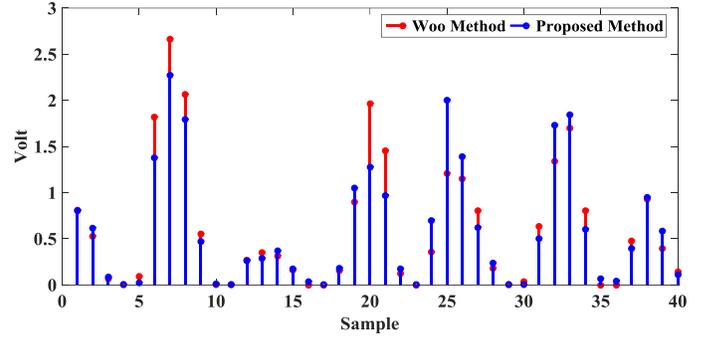

**Fig. 8.** The output ($V_T$) of the proposed model compared to the method of [6] (Woo's model) for the discrete quasi-triangular (Q-triangular) testing data set.

For quantitative evaluation of the precision of the proposed method, a numerical criterion is used. The first criterion is the percentage of Normalized Mean Square Error (NMSE%). The calculated NMSE% between the proposed model and analytical one is 0.435% on average, which indicates that the error of the proposed model is negligible and it can be a proper alternative for the method of [6] (Woo's model). The next argument in evaluating the proposed model is the speed of the algorithm, as compared to that of [6]. Thus, both methods are simulated on an ASUS laptop with an Intel Corei5 processor with 4 GB of RAM using MATLAB software, and the runtimes of the algorithms are measured. Table 1 compares the runtime of the method of [6] ($RT_{Woo}$) with the runtime of the proposed model ($RT_P$) for different datasets, including the chirp, sinusoidal and quasi-triangular signal (Q-Triangular) signals, as well as the mean result. In order to compare the methods, merit called the Speed-Up Ratio (*SUR*) is defined, as

$$SUR = \frac{RT_{Woo}}{RT_P} \quad (25)$$

As shown in Table 1, the mean value of *SUR* over different tests is 313. Hence, on average, the proposed method is 313 times faster than the method of [6], which shows the effectiveness of the proposed technique.

**Table 1:** Comparison of the Woo's model with the proposed model.

| *Signal* | *NMSE%* | $RT_P$ (s) | $RT_{Woo}$ (s) | *SUR* |
|---|---|---|---|---|
| Chirp | 0.0556% | 1.2704e-04 | 0.0472 | 371.68 |
| Sinusoidal | 0.54% | 1.1857e-04 | 0.0397 | 335.13 |
| Q-Triangular | 0.71% | 1.4821e-04 | 0.0342 | 230.96 |
| Mean | 0.4352% | 1.3127e-04 | 0.0403 | 312.59 |

To analyze the error distribution of the proposed method, the following criterion is introduced to compare the proposed method and the analytical method proposed in [6]:

$$D_{VT} = VT_{Woo} - VT_P \quad (26)$$

where $VT_P$ and $VT_{Woo}$ are the calculated threshold voltage using the proposed method and the method of [6], respectively. The normalized histograms of $D_{VT}$ for the three mentioned datasets have been depicted in Figs. 9-11. As show, the mean of all three graphs are almost zero, which proves that the proposed model can be a suitable alternative for the method of [6] (Woo's model). To show the numerical statistics of the $D_{VT}$, its standard deviation ($\sigma_{D_{VT}}$) and the absolute value of the mean ($|\mu_{D_{VT}}|$) have been shown in Table 2. Ideally, both $\sigma_{D_{VT}}$ and $|\mu_{D_{VT}}|$ should be zero. The results that are reported in Table 2 show that, on average, there is 0.01936 volt difference between the mean value of the calculated threshold voltage of the proposed method and the method of [6] with a 0.1636 volt deviation. This result can assure us that the proposed method is accurate and reliable and it can be an appropriate alternative for the model proposed in [6].

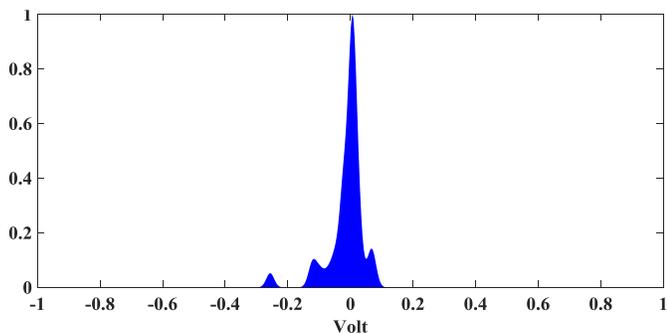

**Fig. 9.** Normalized histogram of $D_{VT}$ for the discrete Chirp testing data set

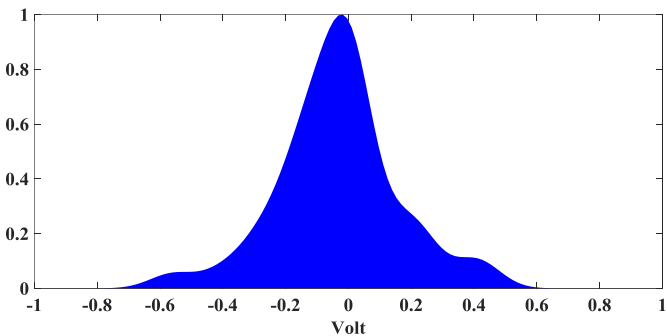

**Fig. 10.** Normalized histogram of $D_{VT}$ for the discrete sinusoidal testing data set

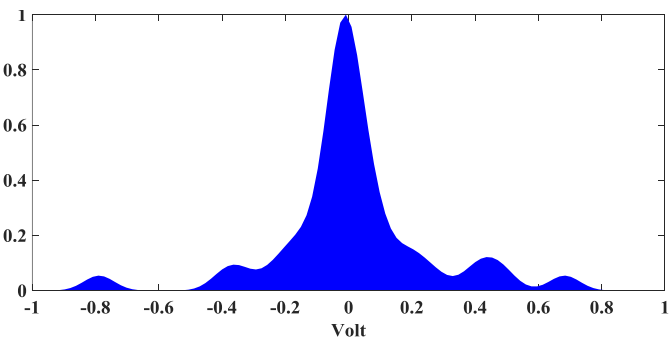

**Fig. 11.** Normalized histogram of $D_{VT}$ for the discrete quasi triangular (Q-triangular) testing data set

**Table 2:** Standard deviation ($\sigma$) and the absolute value of the mean ($|\mu|$) of $D_{VT}$.

| Signal | $|\mu_{D_{VT}}|$ (Volt) | $\sigma_{D_{VT}}$ (Volt) |
|---|---|---|
| Chirp | 0.0122 | 0.0575 |
| Sinusoidal | 0.0348 | 0.191 |
| Q-Triangular | 0.0111 | 0.2423 |
| Mean | 0.01936 | 0.1636 |

## V. CONCLUSION

A method for calculation of the threshold voltage of Junctionless Double-Gate MOSFETs has been proposed using a MISO Nonlinear Autoregressive (N-AR) model. This method is equipped to Gram-Schmidt orthogonalization approach, and it can reduce the high computational load, due to considering the effect of DIBL in analytical methods. Therefore, the proposed method can effectively increase the speed of integrated circuits simulation tools. Numerically, it is shown that, on average, the proposed method is almost 313 times faster than the similar analytical model, with a good accuracy. The proposed method is not only limited to the JL-DG-MOSFET modeling, and it can be used to model other devices.

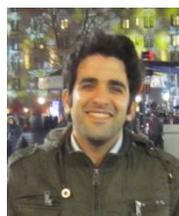

**Mohsen Annabestani** received the M.S. of Biomedical engineering in 2014 and a Ph.D. of Electrical engineering in 2018 both from the Ferdowsi University of Mashhad, Mashhad, Iran. He also pursued the part of his PhD researches on Fabrication and Improvement of Ionic soft actuators and Soft Biorobotics at Harvard university and Harvard-MIT division of health science and technology, Cambridge, MA, USA for six months in 2016. His research interests have been categorized in two parts, the first part is related to computer science and includes artificial Intelligence, fuzzy control, neural networks, deep and hierarchical networks, dynamic system identification and modeling, quantum computation etc. And the second part is related to design, fabrication, modeling and application of soft sensors, soft actuators and smart materials including Electroactive Polymers (EAPs), Shape Memory Alloys (SMAs), especially Ionic Polymer-Metal Composites (IPMCs) etc.

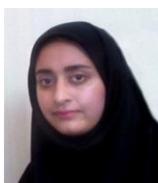

**Mahshid Nasserain** received the B.S. and M.S. degrees from Ferdowsi University of Mashhad, Mashhad, in 2010, 2012, respectively, all in electrical engineering. She is currently a visiting Ph.D. student at Aarhus University.

She is currently pursuing the Ph.D. degree at the same university. Her research interests include design of low-power and high-performance mixed-signal/digital CMOS circuits/systems and also biomedical sensors, and instrumentation.

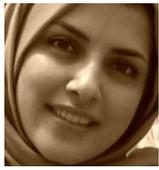

**Fatemeh Hasanzadeh** received the B.S. of Electrical Engineering in 2012 and a M.S. of Biomedical engineering in 2015 both from the Ferdowsi University of Mashhad, Mashhad, Iran. She is currently pursuing the Ph.D. degree in Biomedical Engineering at the K.N.Toosi University of Technology, Tehran, Iran. Her research interests include neuroscience, biological system modeling, dynamic system identification, Brain networks, Brain connectivity, Biomedical signal processing (especially EEG signals) etc.

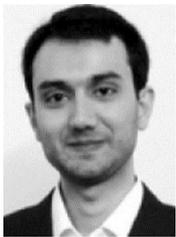

**Mohammad Taherzadeh-Sani** received the B.Sc. degree from the Ferdowsi University of Mashhad, Iran, in 2001, the M.Sc. degree from the University of Tehran, Iran, in 2004, and the Ph.D. degree from McGill University, Montreal, Canada, in 2011. He was a recipient of a J. W. McConnell Memorial Fellowship from McGill University in 2007 and 2008 for his doctoral research, and a Post-Doctoral Fellowship from the Le Fonds Québécois de la Recherche sur la Nature et les Technologies for 2012 and 2013 (declined). In 2012, he joined Ferdowsi University of Mashhad as an Assistant Professor. He authored several journal publications in distinguished journals (e.g., JSSC, TCAS-I, TCAS-II, and T-VLSI) and many papers in different conferences (e.g., ESSCIRC, A-SSCC, ICCAD, and ISCAS). His research interests focuses on biomedical circuits and systems, high-quality an high speed data converters, and radio frequency integrated circuits. He has different fabricated ICs and publications on these subjects. He fabricated several integrated circuits in various technologies from 65- to 180-nm CMOS.

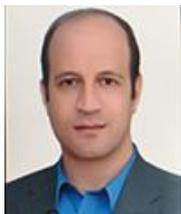

**Alireza Hassanzadeh** received his M.Sc. and Ph.D. degrees in electrical engineering from the University of Alabama in Huntsville, USA, in 2008 and 2011 respectively. He was a member of the Nano and Micro Device Center (NMDC) at the University of Alabama in Huntsville, and Phi-Kappa-Phi, and Eta-Kappa-Nu honor societies. His research interests include low power analog and digital integrated circuits and liquid crystal micro-sensors. He is currently with the department of electrical engineering, Shahid Beheshti University, Tehran, Iran.